%% file: main_2025_03_10.tex
\documentclass{ifacconf}
\usepackage{graphicx}      
\usepackage{amsmath}     
\usepackage{natbib}
\usepackage{etoolbox}  

\usepackage{tikz}
\usetikzlibrary{angles}
\input{PackagesCommands}
\begin{document}
\begin{frontmatter}
%


\title{Adaptive Control of Dual-Rotor Rotational System with Unknown Geometry and Unknown Inertia}


\author{Mohammad Mirtaba, Jhon Manuel Portella Delgado, and Ankit Goel}
\address{Department of Mechanical Engineering, \\ University of Maryland, Baltimore County, \\ 1000 Hilltop Circle, Baltimore, MD 21250.  \\ (e-mail: m.mirtaba, jportella, ankgoel@umbc.edu).}

\begin{abstract}                

This paper develops an input-output feedback linearization-based adaptive controller to stabilize and regulate a dual-rotor rotational system (DRRS), whose inertial properties as well as the geometric configuration of rotors are unknown. 
First, the equations of motion governing the dynamics of DRRS are derived using the Newton-Euler approach. 
Next, an input-output feedback linearization technique is used to linearize the dynamics from the rotor speeds to the angular position of the system. 
A finite-time convergent estimator, based on the portion of the DRRS dynamics, is used to update the required parameters in the controller.
Finally, the proposed controller is validated in both step and harmonic command-following problems, and the robustness of the controller to the system's parameters is demonstrated.
\end{abstract}

\begin{keyword}
adaptive control, dual-rotor rotational system, input-output linearization
\end{keyword}

\end{frontmatter}

\maketitle

\begin{abstract}
 
\end{abstract}

\section{Introduction}
\label{sec:introduction}

Various engineering systems, such as multicopters and humanoid robots, consist of several rotational subsystems that must be precisely controlled to obtain the desired performance. 
In general, control systems for such large-scale systems consist of several hierarchical control systems designed to obtain the desired dynamic behavior of each subsystem. 
%
This paper is focused on designing a control system for such a representative subsystem. 
In particular, this paper focuses on controlling a dual-rotor rotational system (DRRS) with unknown physical properties and geometry. 
A DRRS is an abstraction of various dynamic systems in mechanical and aerospace engineering, such as helicopters, multicopters, vertical take-off and landing systems, etc., and as such, it additionally serves as an excellent testbed for benchmarking novel control and estimation techniques.



A dual-rotor rotational system has been studied extensively since the introduction of Quanser Aero, which was introduced in 2016.
A detailed model of the Quanser Aero with frictional and centripetal forces is derived in \cite{dyvik2023modeling}.
Linearization-based control techniques are investigated in \cite{frasik2018practical, fellag20242, Investigating,pereda2024modeling}, however, linearization-based techniques are only valid in the local state space near the linearization state and thus do not provide stability guarantees. 
%
Nonlinear control techniques such as adaptive backstepping control are investigated in \cite{schlanbusch2019adaptive, schlanbusch2024adaptive}, 
adaptive sliding mode control is investigated in \cite{adaptive-slide}, and a nonlinear controller with feedforward estimates of the input-multiplicative in lieu of adaptive parameters is investigated in \cite{back2}
A data-driven strategy combined with sliding mode control is investigated in \cite{baciu2024model}, however, the paper is focused on only the pitch control of the rotational system. 
%
Alternatively, a data-driven approach that first uses input-output data to identify a linear model and then constructs a control for the linear model is explored in \cite{id}.
While many recent studies have developed adaptive controllers for dual-rotor systems, to the author's knowledge, no previous research has addressed the problem without knowing the rotor configuration and the mapping that relates rotor speeds to forces and torques.

The DRRS considered in this paper consists of a rigid body with unknown inertial properties that rotates about two noncollinear axes independently and two noncoplanar rotors that can generate forces.
%
In this work, we assume that the rotors are mounted at unknown angles and thus generate forces and moments along unknown directions.
Furthermore, we assume that rotor coefficients relating the rotor speed to the generated force and torque are unknown.  
To design the control system, a feedback linearization technique described in \cite{portella2024adaptive,portella2024circumventing} is used to linearize the input-output dynamics.
Next, an output tracking controller is designed using the classical linear quadratic technique.
Finally, an estimation system with finite-time convergence properties is designed to estimate the parameters required in the control system.



The paper is organized as follows. 
The equations of motion of the dual-rotor rotational system are derived in detail using the Newton-Euler dynamics in Section \ref{sec:platform}.
An adaptive input-output feedback linearization control is developed for the DRRS in Section \ref{sec:control}.
The adaptive control system's application to follow step and harmonic commands is demonstrated in Section \ref{sec:simulation}.
Finally, the paper concludes with a summary and discussion of future directions in Section \ref{sec:conc}.


\section{Dual-Rotor Rotational System}
\label{sec:platform}

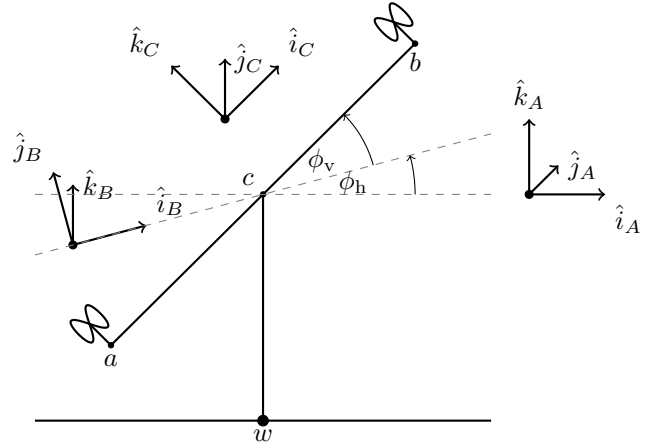
\begin{figure}[h]
    \centering
    \begin{tikzpicture}
    \coordinate (A) at (0,0);    
    \coordinate (B) at (6,0);    
    \coordinate (C) at (3,0);    
    \coordinate (D) at (3,3);    

    \coordinate (E1) at ($(D) + (-2,-2)$);  
    \coordinate (E2) at ($(D) + (2,2)$);    

    \pgfmathsetmacro\xperp{0.4*cos(45)}  
    \pgfmathsetmacro\yperp{0.4*sin(45)}

    \draw[thick] (A) -- (B);
    
    \draw[thick] (C) -- (D);
    
    \draw[thick] (E1) -- (E2);
    
    \filldraw[black] (E1) circle (1pt) node[below ] {$a$}; 
    \filldraw[black] (E2) circle (1pt) node[below ] {$b$}; 
    \filldraw[black] (D) circle (1pt) node[above left] {$c$}; 
    \filldraw[black] (C) circle (2pt) node[below] {$w$}; 
    
    \coordinate (P1) at ($(E1) + (-\yperp,\xperp)$);
    \coordinate (P2) at ($(E2) + (-\yperp,\xperp)$);

    \draw[thick] (E1) -- (P1);
    \draw[thick] (E2) -- (P2);

    \begin{scope}[rotate around={45:(P1)}]
        \draw[thick, domain=0:360, samples=50, scale=0.2, shift={(P1)}] plot ({cos(\x)}, {sin(2*\x)});
    \end{scope}

    \begin{scope}[rotate around={45:(P2)}]
        \draw[thick, domain=0:360, samples=50, scale=0.2, shift={(P2)}] plot ({cos(\x)}, {sin(2*\x)});
    \end{scope}

\coordinate (F) at ($(E2) + (-2.5,-1)$);  


\coordinate (Fx) at ($(F) + ({1*cos(45)}, {1*sin(45})$);  
\coordinate (Fy) at ($(F) + ({-1*sin(45)}, {1*cos(45)})$); 
\coordinate (Fz) at ($(F) + (0,0.8)$);  

\draw[thick, ->] (F) -- (Fx) node[above right] {$\hat i_C$};
\draw[thick, ->] (F) -- (Fy) node[above left] {$\hat k_C$};
\draw[thick, ->] (F) -- (Fz) node[right] {$\hat j_C$};

\filldraw[black] (F) circle (1.5pt);

\coordinate (F) at ($(E2) + (1.5,-2,0)$);

\coordinate (Fx) at ($(F) + (1,0,0)$);  
\coordinate (Fy) at ($(F) + (0,1,0)$);  
\coordinate (Fz) at ($(F) + (0,0,-1)$);  

\draw[thick, ->] (F) -- (Fx) node[below right] {$\hat i_A$};
\draw[thick, ->] (F) -- (Fy) node[above] {$\hat k_A$};
\draw[thick, ->] (F) -- (Fz) node[right] {$\hat j_A$};

\filldraw[black] (F) circle (1.5pt);

\coordinate (G) at ($(D) + (-2.5,{-2.5*tan(15)})$);  


\coordinate (Gx) at ($(G) + ({1*cos(15)}, {1*sin(15)})$);  
\coordinate (Gy) at ($(G) + ({-1*sin(15)}, {1*cos(15)})$); 
\coordinate (Gz) at ($(G) + (0,0.8)$);  

\draw[thick, ->] (G) -- (Gx) node[above right] {$\hat i_B$};
\draw[thick, ->] (G) -- (Gy) node[above left] {$\hat j_B$};
\draw[thick, ->] (G) -- (Gz) node[right] {$\hat k_B$};

\filldraw[black] (G) circle (1.5pt);

    \coordinate (T1) at ($(D) + (-3,0)$); 
    \coordinate (T2) at ($(D) + (3,0)$);  
    \coordinate (T3) at ($(D) + (-3,{-3*tan(15)})$); 
    \coordinate (T4) at ($(D) + (3,{3*tan(15)})$);   

    \draw[gray, dashed, thin] (T1) -- (T2); 
    \draw[gray, dashed, thin] (T3) -- (T4); 

    \draw pic [draw=black, angle radius=20mm, pic text=$\phi_\mathrm{h}$,->] {angle = T2--D--T4};
    \draw pic [draw=black, angle radius=15mm,pic text=$\phi_\mathrm{v}$,->] {angle = T4--D--E2};
\end{tikzpicture}

    \caption{Dual-rotor rotational system.}
    \label{fig:2DRotatingPlatform}
\end{figure}

A dual-rotor rotational system, shown in Figure \ref{fig:2DRotatingPlatform}, consists of a vertical rigid body $wc,$ which is fixed to the ground, and the rigid body $ab$, which is connected to the vertical body at $c$ with a ball joint.
The body $ab$ can rotate at the ball joint in the horizontal plane (with angle $\phi_\rmh)$ as well as the vertical plane (with angle $\phi_\rmv$).
Two rotors are mounted at $a$ and $b$ whose axis lies in the plane orthogonal to $ab.$

As shown in Figure \ref{fig:2DRotatingPlatform}, 
let $\rm F_A$ be a frame fixed to the ground,
let $\rm F_B$ be defined such that $\khat B$ is aligned with the vertical section $oc$ and $\ihat B$ is aligned with the projection of $cb$ on the $\ihat A-\jhat A$ plane,
and let $\rm F_C$ be defined such that $\ihat C$ is along $cb$ and $\khat C$ lies in the $\ihat B-\khat B$ plane.
Note that $\rm F_C$ is fixed to the rigid arm $ab,$ denoted by $\SB.$
The frames are thus related by
\begin{align}
    \rmF_\rmA 
        \mathop{\longrightarrow}^{\phi_\rmh}_{3} 
            \rmF_\rmB
        \mathop{\longrightarrow}^{\phi_\rmv}_{2} 
            \rmF_\rmC,
    \label{eq:frame_relation}
\end{align}
where $\phi_\rmh$ is the Euler angle about $\khat A = \khat B$ and 
$\phi_\rmv$ is the Euler angle about $\jhat B = \jhat C.$
Note that $\phi_\rmh$ is the rotation angle in the horizontal plane, and $\phi_\rmv$ is the rotation angle in the vertical plane.
Next, it follows from \eqref{eq:frame_relation} that the angular velocity $\vect \omega_{\rm C/A}$ of the arm $\SB$ relative to the inertial frame $\rm F_A$ is
\begin{align}
    \vect \omega_{\rm C/A} 
        &=
            \dot \phi_\rmv \jhat \rmC +
            \dot \phi_\rmh \khat \rmB
        \nn \\
        &=
            \dot \phi_\rmv \jhat \rmC +
            \dot \phi_\rmh (-\sin \phi_\rmv\ihat \rmC + \cos \phi_\rmv \khat \rmC)
        \nn \\
        &=
            -\sin \phi_\rmv \dot \phi_\rmh \ihat \rmC +
            \dot \phi_\rmv \jhat \rmC +
            \cos \phi_\rmv \dot \phi_\rmh \khat \rmC,
\end{align}
and thus the angular acceleration $\framedot{C}{\vect \omega}_{\rm C/A}$ of the arm $\SB$ relative to the inertial frame $\rm F_A$ relative to the body-fixed frame $\rm F_C$ is
\begin{align}
    \framedot{C}{\vect \omega}_{\rm C/A} 
        &=
            -(\sin \phi_\rmv \ddot \phi_\rmh 
            +\cos \phi_\rmv \dot \phi_\rmh) \ihat \rmC
            \nn \\ &\quad +
            \ddot \phi_\rmv \jhat \rmC +
            (\cos \phi_\rmv \ddot \phi_\rmh - \sin \phi_\rmv \dot \phi_\rmh ) \khat \rmC.
\end{align}

Since the $\rm F_C$ is a principal axis frame, it follows that the physical inertia matrix can be written as  $\tarrow J_{\SB/c} = J_1 \ihat \rmC \ihat \rmC' + J_2 \jhat \rmC \jhat \rmC' + J_3 \khat \rmC \khat \rmC'.$
Therfore, 
\begin{gather}
    \tarrow J_{\SB/c} \framedot{C}{\vect \omega}_{\rm C/A} 
        =
            -J_1 (\sin \phi_\rmv \ddot \phi_\rmh +\cos \phi_\rmv \dot \phi_\rmh) \ihat \rmC
            +
            J_2 \ddot \phi_\rmv \jhat \rmC
            \nn \\ \quad +
            J_3 (\cos \phi_\rmv \ddot \phi_\rmh - \sin \phi_\rmv \dot \phi_\rmh ) \khat \rmC,
            \\
    \vect \omega_{\rm C/A}  \times \tarrow J_{\SB/c} \vect \omega_{\rm C/A} 
        =
            (-J_2 
            +
            J_1  ) \sin \phi_\rmv \dot \phi_\rmv \dot \phi_\rmh \khat \rmC
            \nn \\ \quad +
            (J_3 
            -J_1 ) \sin \phi_\rmv \cos \phi_\rmv \dot \phi_\rmh ^2 \jhat \rmC
            +
            (J_3 
            -
            J_2) \cos \phi_\rmv \dot \phi_\rmh \dot \phi_\rmv \ihat \rmC.
\end{gather}

Next, define frames $\rm F_{D_a}$ and $\rm F_{D_b}$ fixed to the motors at $a$ and $b,$ respectively, such that 
\begin{align}
    \rmF_\rmC 
        &\mathop{\longrightarrow}^{\beta_a}_{1} 
            \rm F_{D_a}, \quad 
    \rmF_\rmC 
        \mathop{\longrightarrow}^{\beta_b}_{1} 
            \rm F_{D_b},
\end{align}
where $\beta_a$ and $\beta_b$ are the angles of the motor axis around the arm $\SB.$
Assume that the reaction forces $f_a$ and $f_b$ and the reaction torques $\tau_a$ and $\tau_b$ due to the motors at $a$ and $b$ are parameterized as 
\begin{gather}
    f_a     \isdef k_f p(\omega_a), \quad 
    f_b     \isdef k_f p(\omega_b), \\
    \tau_a  \isdef k_\tau p(\omega_a), \quad 
    \tau_b  \isdef -k_\tau p(\omega_b), 
\end{gather}
where the function $p(\omega) \isdef \omega | \omega|.$

Next, note that 
\begin{align}
    \vect M_{\SB/c}
        &=
            \vect r_{a/c} \times \vect f_a \jhat{D_a} + \tau_a \khat \rmC
            +
            \vect r_{b/c} \times \vect f_b + \tau_b \khat \rmC + \tau_\rmr \ihat \rmC
        \nn 
        \\
        &=
            \tau_\rmr \ihat \rmC
            +
            \ell 
            (
                 f_a \cos \beta_a - f_b \cos \beta_b
            ) \jhat{C}
            \neweqline +
            \ell
            (
                f_a \sin \beta_a -  f_b  \sin \beta_b 
            )
            \khat C 
            \neweqline
            +
            (- \tau_a\sin \beta_a - \tau_b\sin \beta_b ) \jhat C
            \neweqline 
            +
            ( \tau_a\cos \beta_a  + \tau_b \cos \beta_b )    \khat C
\end{align}
and thus the Euler's equation implies that 
\begin{gather}
    J_2 \ddot \phi_\rmv + (J_3 
            -J_1 ) \sin \phi_\rmv \cos \phi_\rmv \dot \phi_\rmh ^2
        =
            M_\rmv, 
    \label{eq:thetaddot}
    \\
    J_3 (\cos \phi_\rmv \ddot \phi_\rmh - \sin \phi_\rmv \dot \phi_\rmh )
        +
    (-J_2 
            +
            J_1  ) \sin \phi_\rmv \dot \phi_\rmv \dot \phi_\rmh
        = 
            M_\rmh,
    \label{eq:phiddot}
\end{gather}
where
\begin{align}
    M_\rmv
        &\isdef
            \ell 
            (
                 k_f p(\omega_a)  \cos \beta_a - k_f p(\omega_b) \cos \beta_b
            )
            \neweqline +
            (- k_\tau p(\omega_a) \sin \beta_a + k_\tau p(\omega_b) \sin \beta_b )
    , \\
    M_\rmh
        &\isdef 
            \ell
            (
                k_f p(\omega_a) \sin \beta_a -  k_f p(\omega_b)  \sin \beta_b 
            )
            \neweqline +
            ( k_\tau p(\omega_a) \cos \beta_a  - k_\tau p(\omega_b) \cos \beta_b ).
\end{align} 
which can be concisely written using a control allocation matrix $\SC$ as 
\begin{align}
    \begin{bmatrix}
        M_\rmv\\
        M_\rmh
    \end{bmatrix} &=
    \SC
    \begin{bmatrix}
        p(\omega_a) \\
        p(\omega_b)
    \end{bmatrix},
\end{align}
where the components of the control allocation matrix $\SC$ are
\begin{align}
    \SC(1,1) &= {k_\rmf \ell \cos \beta_a - k_\tau \sin \beta_a} , \\
    \SC(1,2) &= {-k_\rmf \ell \cos \beta_b + k_\tau \sin \beta_b}, \\
    \SC(2,1) &= {k_\rmf \ell \sin \beta_a + k_\tau \cos \beta_a}, \\
    \SC(2,2) &= {-k_\rmf \ell \sin \beta_b - k_\tau \cos \beta_b}.    
\end{align}


\begin{remark}

Let $\beta_a = \beta_b = 0.$ Then, 
\begin{align}
    \matl
        M_\rmv\\
        M_\rmh
    \matr 
    &=
    \matl
        k_f \ell   & -k_f \ell  \\
        k_\tau  &  - k_\tau 
    \matr 
    \matl
        p(\omega_a) \\
        p(\omega_b)
    \matr.
\end{align}
In this configuration, arbitrary values of $M_\rmv$ and $M_\rmh$ can not be generated since the control allocation matrix is rank-deficient. 
\end{remark}

\begin{remark}
    Let $\beta_a = 0$ and $\beta_b = \pi/2.$ Then, 
\begin{align}
    \matl
        M_\rmv\\
        M_\rmh
    \matr 
        \isdef
    \matl
        k_f \ell   & k_\tau  \\
        k_\tau  &  - k_f \ell
    \matr 
    \matl
        p(\omega_a) \\
        p(\omega_b)
    \matr.
\end{align}
In this configuration, arbitrary values of $M_\rmv$ and $M_\rmh$ can be generated since the control allocation matrix is not rank-deficient. 
\end{remark}

\begin{remark}
    Note that the determinant of the control allocation matrix is 
    $(\ell^2 k_f^2 + k_\tau^2) \sin (\beta_a- \beta_b),$ which implies that arbitrary values of $M_\rmv$ and $M_\rmh$ can be generated if $\beta_a - \beta_b \neq n \pi,$ where $n \in \BBN.$
\end{remark}


\section{Adaptive Control System}
\label{sec:control}
This section presents the adaptive controller to track the angular position commands. 
In particular, an input-output linearizing controller is coupled with a finite-time estimation system to construct the adaptive controller. 

To design an input-output linearizing controller, we first write the equations of motion in the strict-feedback form, as shown below. 
Defining
\begin{align}
    \xi_1 
        \isdef
            \matl
                \phi_\rmv\\
                \phi_\rmh
            \matr,
    \quad 
    \xi_2 \isdef \matl
        \dot \phi_\rmv\\
        \dot \phi_\rmh
    \matr,
    \quad 
    \Omega \isdef \matl
        p(\omega_a) \\
        p(\omega_b)
    \matr,
\end{align}
it follows from the equations of motion \eqref{eq:thetaddot}, \eqref{eq:phiddot} that
\begin{align}
    \dot \xi_1 &= \xi_2,
    \label{eq:xi1_dot}
    \\
    \dot \xi_2 &= \SF_1(\xi_1,\xi_2) + \SF_2(\xi_1,\xi_2)\Theta_1 + \SG(\xi_1) \Theta_2 \Omega,
    \label{eq:xi2_dot}
\end{align}
where
\begin{align}
    \SF_1(\xi_1,\xi_2)
        &\isdef
            \matl
                0\\
                \dfrac{\sin(\phi_\rmv)\dot \phi_\rmh}{\cos(\phi_\rmv)}
            \matr,
    \\
    \SF_2(\xi_1,\xi_2)
        &\isdef
            \matl
                \sin(\phi_\rmv)\cos(\phi_\rmv)\dot \phi_\rmh^2 & 0 
                \\
                0 & \dfrac{\sin(\phi_\rmv)}{\cos(\phi_\rmv)}\dot \phi_\rmv\dot \phi_\rmh
            \matr
    \\
    \SG(\xi_1)
        &\isdef
        \matl
            1 & 0\\
            0 & \dfrac{1}{\cos{\phi_\rmv}}
        \matr,
    \\
    \Theta_1
        &\isdef
            \matl
                \dfrac{J_1-J_3}{J_2}
                \\
                \dfrac{J_2-J_1}{J_3}
            \matr,
\end{align}
and $\Theta_2 \in \BBR^{2\times 2}$ whose entries are
{\small
\begin{align}
    \Theta_2(1,1) &= \dfrac{k_\rmf \ell \cos \beta_a - k_\tau \sin \beta_a}{J_2} , \\
    \Theta_2(1,2) &= \dfrac{-k_\rmf \ell \cos \beta_b + k_\tau \sin \beta_b}{J_2}, \\
    \Theta_2(2,1) &= \dfrac{k_\rmf \ell \sin \beta_a + k_\tau \cos \beta_a}{J_3}, \\
    \Theta_2(2,2) &= \dfrac{-k_\rmf \ell \sin \beta_b - k_\tau \cos \beta_b}{J_3}.    
\end{align}}%
Note that $\Theta_1$ and $\Theta_2$ are constant parameters that depend only on the system's physical properties.
Furthermore, the functions $\SF_1, \SF_2,$ and $\SG$ are undefined at $\phi_\rmv = \pi/2.$
This is due to the choice of Euler angles to parameterize the orientation of the DRRS.




Next, to design the input-output linearizing controller for the system \eqref{eq:xi1_dot}, \eqref{eq:xi2_dot}, we write the system as 
\begin{align}
    \dot x &= f(x) + g(x) u, \\
    y &= h(x),
\end{align}
where 
\begin{align}
    x
        =
            \matl 
                \xi_1 \\
                \xi_2 
            \matr \in \BBR^4, 
\end{align}
and 
\begin{align}
    f(x)
        &\isdef 
            \matl 
                \xi_2 \\
                \SF_1(\xi_1,\xi_2) + \SF_2(\xi_1,\xi_2)\Theta_1
            \matr , \
    g(x)
        \isdef
            \matl 
                0_{2 \times 2} \\
                \SG(\xi_1) \Theta_2
            \matr.
\end{align}
Letting $\xi_1$ be the output of the system implies that
\begin{align}
    h(x) = \xi_1 \in \BBR^2.
\end{align}

Note that the relative degree $\rho_1$ of output $y_1$ with respect to the input $\omega_\rma$ is $2$
and the relative degree $\rho_2$ of output $y_2$ with respect to the input $\omega_\rmb$ is $2.$
Since $\rho \isdef \rho_1 + \rho_2 = 4 = l_x,$ it follows that the system does not have zero dynamics. 

\subsection{Input-Output Linearizing Control}

As shown in \cite{portella2024circumventing}, the input-output linearizing controller is given by 
\begin{align}
    u(x) = \beta(x)^{-1}(-\alpha(x) + v),
    \label{eq:IOL_control_law}
\end{align}
where 
\begin{align}
    \alpha(x)
        &\isdef
            \matl 
                L_f^{2} h_1(x) \\[0.5em]
                L_f^{2} h_{2}(x)
            \matr
        =
        \SF_1(\xi_1,\xi_2) + \SF_2(\xi_1,\xi_2)\Theta_1 
        \in \BBR^{2}, 
    \\
    \beta(x)
        &\isdef 
            \matl 
                L_g L_f h_1(x) \\[0.5em]
                L_g L_f h_{2} (x)
            \matr
            = \SG(\xi_1) \Theta_2
            \in \BBR^{2 \times 2}.
\end{align}

With the controller \eqref{eq:IOL_control_law}, it follows that 
\begin{align}
    \dot x
        &=
            A_\rmc x 
            +
            B_\rmc  v, 
    \label{eq:IOlinearizedsystem_x}
    \\
    y
        &=
            C_\rmc x,
    \label{eq:IOlinearizedsystem_y}
\end{align}
where 
\begin{align}
    A_\rmc &\isdef \begin{bmatrix}
        0 & 0 & 1 & 0\\
        0 & 0 & 0 & 1\\
        0 & 0 & 0 & 0\\
        0 & 0 & 0 & 0
    \end{bmatrix}, 
    \ 
    B_\rmc \isdef \begin{bmatrix}
        0 & 0\\
        0 & 0\\
        1 & 0\\
        0 & 1
    \end{bmatrix},
    \
    C_\rmc \isdef \begin{bmatrix}
        1 & 0 & 0 & 0\\
        0 & 1 & 0 & 0
    \end{bmatrix}.
\end{align}

\subsubsection{Constant Commands.}
To track constant commands, as shown in \cite{portella2024adaptive}, the control law 
\begin{align}
    v = k_x x + k_q q, 
    \label{eq:FSFI}
\end{align}
where $q$ is the integrated output error and satisfies
\begin{align}
    \dot q = r - y,
\end{align}
where $r$ is the commanded reference to be tracked and $y$ is the output of the system, ensure that $\lim_{t \to \infty} r-y = 0.$  

\subsubsection{Time-varying Commands.}
To track time-varying commands, as shown in \cite{portella2024adaptive}, the control law 
\begin{align}
    v = K (x - x_\rmd) + B_\rmc^T \dot x_\rmd,
\end{align}
where $K$ is chosen such that $A_\rmc + B_\rmc K$ is Hurwitz, 
ensure that $\lim_{t \to \infty } \|x-x_\rmd\| = 0.$

%

%


\subsection{Adaptive Augmentation}
The nonlinear controller \eqref{eq:IOL_control_law} requires the parameters $\Theta_1$ and $\Theta_2$ to be implemented. 
Since these parameters are either unknown or uncertain, this section presents a parameter update law that generates the estimates of these unknown parameters online, which can be used to compute the control signals. 

We rewrite \eqref{eq:xi2_dot} as 
\begin{align}
    \dot{\xi}_2 - 
    \SF_1(\xi_1,\xi_2)
        &= 
            \Phi(\xi,\Omega)\Theta,
    \label{eq:regressor}
\end{align}
where
{\small
\begin{align}
    \Phi(\xi,\Omega)
        &\isdef
            \matl
                \sin(x_1)\cos(x_1)x_4^2 & 0
                \\
                0 & \tan(x_1)x_3x_4
                \\
                \Omega_1 & 0
                \\
                \Omega_2 & 0
                \\
                0 & \Omega_1\sec{x_1}
                \\
                0 & \Omega_2\sec{x_1}
            \matr^{\rm T}
            \in \mathbb{R}^{2 \times 6},
\end{align}}
and
{\small
\begin{align}
    \Theta
                &\isdef
                    \matl
                        \dfrac{J_1-J_3}{J_2}
                        \\
                        \dfrac{J_2-J_1}{J_3}
                        \\
                        \dfrac{k_\rmf \ell \cos \beta_a - k_\tau \sin \beta_a}{J_2} 
                        \\ 
                        \dfrac{-k_\rmf \ell \cos \beta_b + k_\tau \sin \beta_b}{J_2}
                        \\
                        \dfrac{k_\rmf \ell \sin \beta_a + k_\tau \cos \beta_a}{J_3} 
                        \\ 
                        \dfrac{-k_\rmf \ell \sin \beta_b - k_\tau \cos \beta_b}{J_3}
                    \matr
                    \in \mathbb{R}^6.
\end{align}}

Since $\xi_1$ and $\Omega$ are known, the signals $\SF_1(\xi_1, \xi_2)$ and $\Phi(\xi_1, \xi_2,\Omega)$ can be directly computed. 
%
%
To compute $\dot{\xi}_2$, we filter \eqref{eq:regressor} with a strictly proper filter $R(s)$ to obtain the linear regressor equation
\begin{align}
    \xi_\rmf
        &=
            \Phi_\rmf\Theta,
\end{align}
where 
\begin{align}
    \xi_\rmf  &\isdef R(s)[\dot{\xi}_2-\SF_1(\xi_1, \xi_2)], \\ 
    \Phi_\rmf &\isdef R(s)[\Phi(\xi,u)].    
\end{align}

For example, leting $R(s)=\dfrac{1}{s+\gamma},$ where $\gamma>0,$ implies that 
\begin{align}
    \xi_\rmf &= \dfrac{s}{s + \gamma} \xi_2 - \dfrac{1}{s + \gamma} \SF_1(\xi_1, \xi_2),    
    \\
    \Phi_\rmf &= \dfrac{1}{s + \gamma} \Phi(\xi,u).  
\end{align}
Note that $x_\rmf$ and $\Phi_\rmf$ can now be computed online using only the measurements of the state $\xi$ and the input $\Omega.$

Finally, consider the estimator
\begin{align}
    \dot{\hat{\Theta}}
        &=
            -c_1 \dfrac{\Xi}{\|\Xi\|_2^{(1-\alpha_1)}}
            -c_2 \dfrac{\Xi}{\|\Xi\|_2^{(1-\alpha_2)}},
    \label{eq:FT_estimator}
\end{align}
where $\Xi \isdef \overline{\Phi}\hat{\Theta} - \overline{\xi} \in \BBR^6,$ and the constants $c_1,c_2 > 0,$ $0 < \alpha_1 < 1,$ and $\alpha_2>1.$
The data matrices $\overline{\xi}$ and $\overline{\Phi}$ satisfy 
\begin{align}
    \dot{\overline{\xi}}
        &=
            -\lambda \overline{\xi} 
            + \Phi_\rmf^{\rm T}\xi_\rmf,
    \quad 
    \dot{\overline{\Phi}}
        =
            -\lambda \overline{\Phi}
            + \Phi_\rmf^{\rm T}\Phi_\rmf,
\end{align}
where $\lambda > 0$ is the exponential forgetting factor.

Finally, the adaptive IOL controller is
\begin{align}
    u(x) 
        &= 
            -\left( \SG(\xi_1) \hat \Theta_2 \right)\inv
            \left( \SF_1(\xi_1,\xi_2) + \SF_2(\xi_1,\xi_2) \hat \Theta_1 +
             v \right) .
    \label{eq:adaptive_IOL_control_law}
\end{align}

\section{Numerical Simulations}
\label{sec:simulation}
This section numerically demonstrates the stabilization and command following the application of the proposed controller. 
To simulate the DRRS, we use the physical parameters shown in Table \ref{tab:parameters}.

The initial angles of the DRRS system $\phi_\rmv(0)$ and $\phi_\rmh(0)$ are set to $0$.
\begin{table}[H]
    \centering
    \renewcommand{\arraystretch}{1.5}
    \begin{tabular}{||c|c||c|c||}
        \hline
        Variable & Value & Variable & Value  \\
        \hline
        \hline
        $J_1$  & 6.25e-4 $\mathrm{kg m^2}$ & $\ell$  & 1 $\mathrm{m}$ \\
        \hline
        $J_2$  & 0.02 $\mathrm{kg m^2}$ & $\beta_a$  & ${\pi}/{4}$ \\
        \hline
        $J_3$  & 0.02 $\mathrm{kg m^2}$ & $\beta_b$  & $-{\pi}/{4}$ \\
        \hline
        $k_\rmf$  & 4e-3 $\mathrm{N s^2/rad^2}$ & $\phi_\rmv(0)$ & 0 \\
        \hline
        $k_\tau$  & 7.5e-4 $\mathrm{N m s^2/rad^2}$ & $\phi_\rmh(0)$ & 0 \\
        \hline
    \end{tabular}
    \caption{Physical parameters of the DRRS.}
    \label{tab:parameters}
\end{table}

\subsection{Constant Command Tracking}
In this example, we consider a step command. 
In particular, for $t \ge 0,$  the reference signal is given by $r(t) = \matl \pi/4 & -\pi/3 \matr^\rmT.$

The adaptive controller is given by \eqref{eq:adaptive_IOL_control_law}, where the internal control signal $v$ given by \eqref{eq:FSFI} is computed using Matlab's LQR
routine, where the LQR weighting matrices $R_1 = 10I_6 $ and $R_2 = I_2.$
%
In the estimator \eqref{eq:FT_estimator}, we set $\gamma = 10^{3}, \lambda = 0.8, c_1 = c_2 = 0.1,$ $\alpha_1  = \alpha_2 = 0.5.$
%
Since $J_2=J_3 >>J_1$ in practice, $\Theta_1 \approx 1$ and $\Theta_2 \approx -1.$
Furthermore, since the control law \eqref{eq:adaptive_IOL_control_law} requires the inverse of $\Theta_2,$ the parameter estimate $\hat \Theta$ is initialized as 
\begin{align}
    \hat \Theta(0) 
        = 
             \matl 1 & -1 & 0.1 & 0 & 0 & 0.1 \matr^\rmT,
\end{align}
which ensures that $\hat \Theta_2 = 0.1I_2,$ and thus $\beta(x)$ is invertible at $t=0.$

\begin{figure}[h]
    \centering
    \includegraphics[width=0.9\columnwidth]{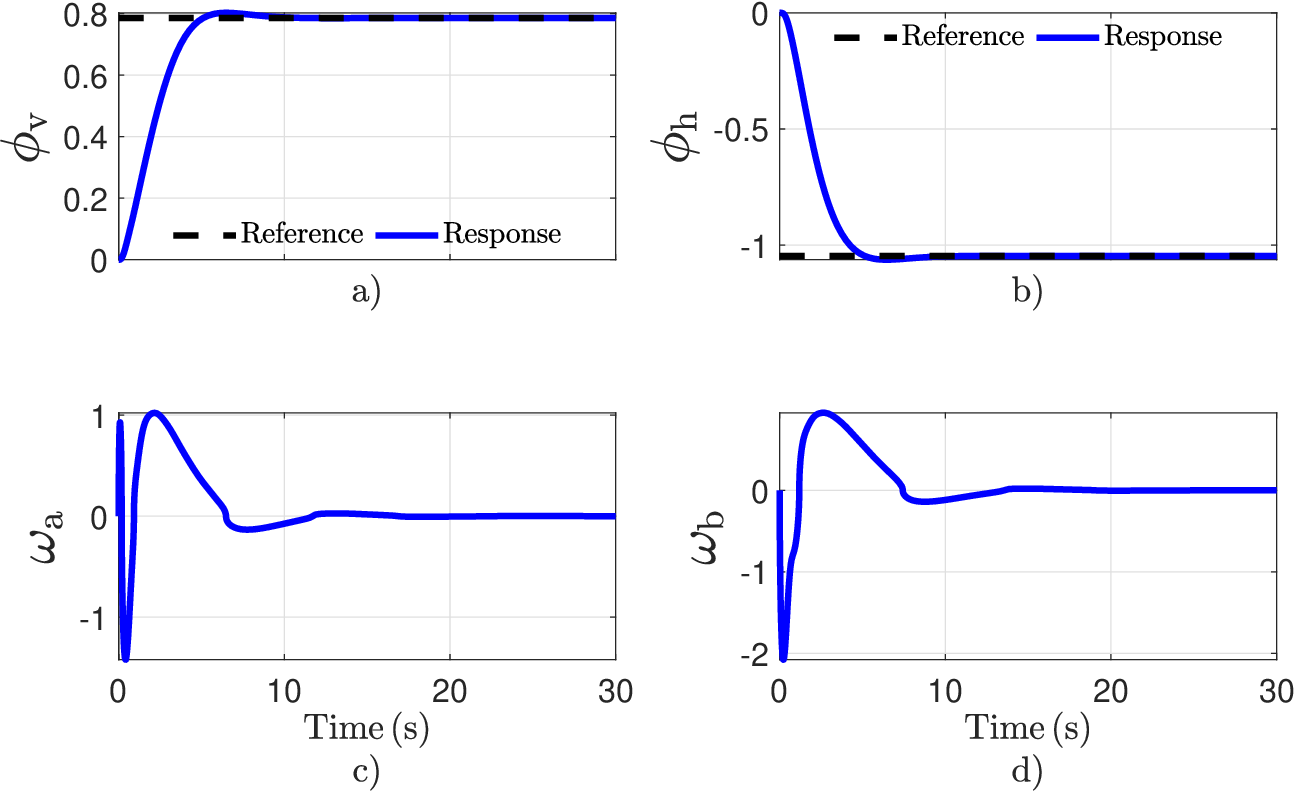}
    \caption{Closed-loop response of the DRRS to a step command with the adaptive input-output linearizing controller \eqref{eq:adaptive_IOL_control_law}.}
    \label{fig:2dof_iol_adapt_control_step}
\end{figure}

Figure \ref{fig:2dof_iol_adapt_control_step} shows the closed-loop response of the DRRS with the adaptive input-output linearizing controller.
a) and b) show the angles $\phi_\rmv$ and $\phi_\rmh,$ respectively and
c) and d) show the propeller speeds $\omega_\rma$ and $\omega_\rmb,$ respectively.




\begin{figure}[h]
    \centering
    \includegraphics[width=0.9\columnwidth]{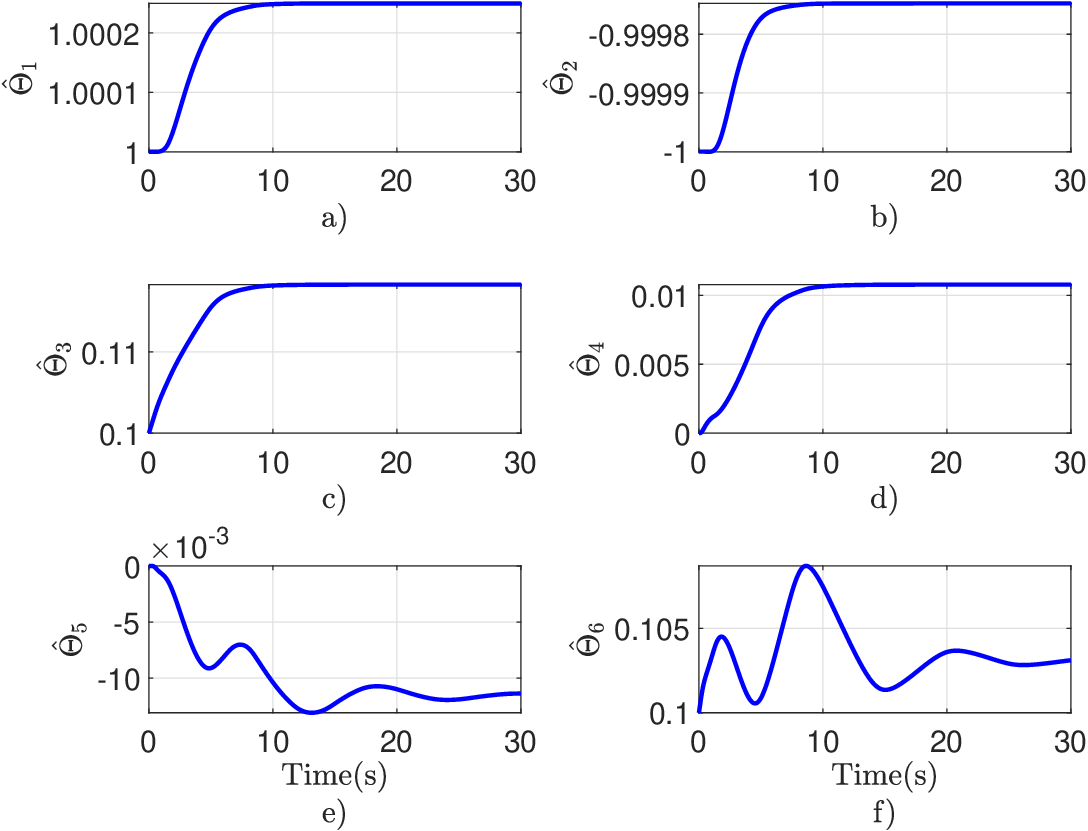}
    \caption{Parameter estimates updated by \eqref{eq:FT_estimator} in the step command-following problem. }
    \label{fig:2dof_iol_adapt_parameters_step}
\end{figure}

Figure \ref{fig:2dof_iol_adapt_parameters_step} shows the parameter estimates updated using the parameter estimator \eqref{eq:FT_estimator}.
Note that the estimates do not necessarily converge to their true values. 

\subsection{Harmonic Command Tracking}
In this example, we consider a harmonic command. 
In particular, we set, for $t \ge 0,$ $r(t) =\frac{\pi}{4} \matl \sin \frac{t}{2} & \cos \frac{t}{2} \matr^\rmT.$
In the controller and the parameter estimator, we use the same setting as constant command tracking.

\begin{figure}[h]
    \centering
    \includegraphics[width=0.9\columnwidth]{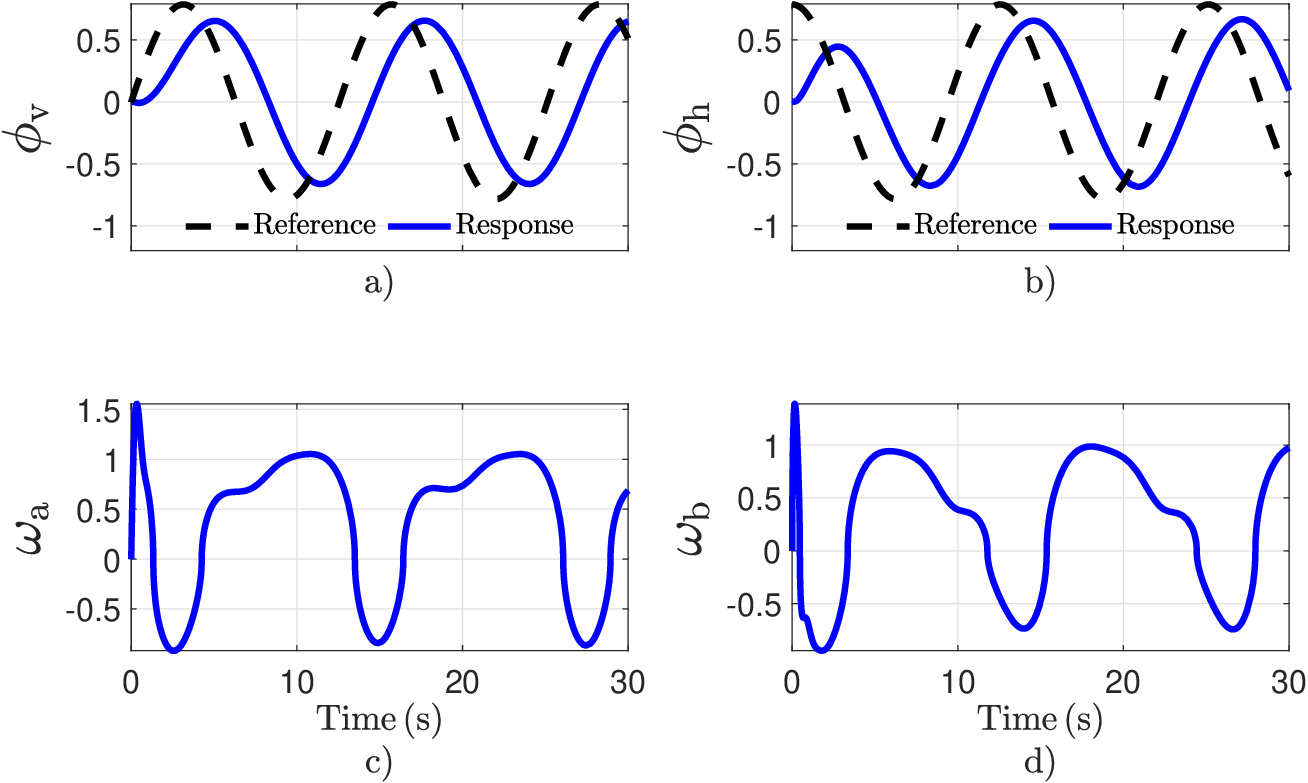}
    \caption{Closed-loop response of the DRRS system to a harmonic command with the adaptive input-output linearizing controller.}
    \label{fig:2dof_iol_adapt_control_sinusoidal}
\end{figure}

Figure \ref{fig:2dof_iol_adapt_control_sinusoidal} shows the closed-loop response of the DRRS with the adaptive IOL controller \eqref{eq:adaptive_IOL_control_law}.
a) and b) show the response of $\phi_\rmv$ and $\phi_\rmh,$ respectively, and
c) and d) show the propeller speeds $\omega_\rma$ and $\omega_\rmb,$ respectively.

\begin{figure}[h]
    \centering
    \includegraphics[width=0.9\columnwidth]{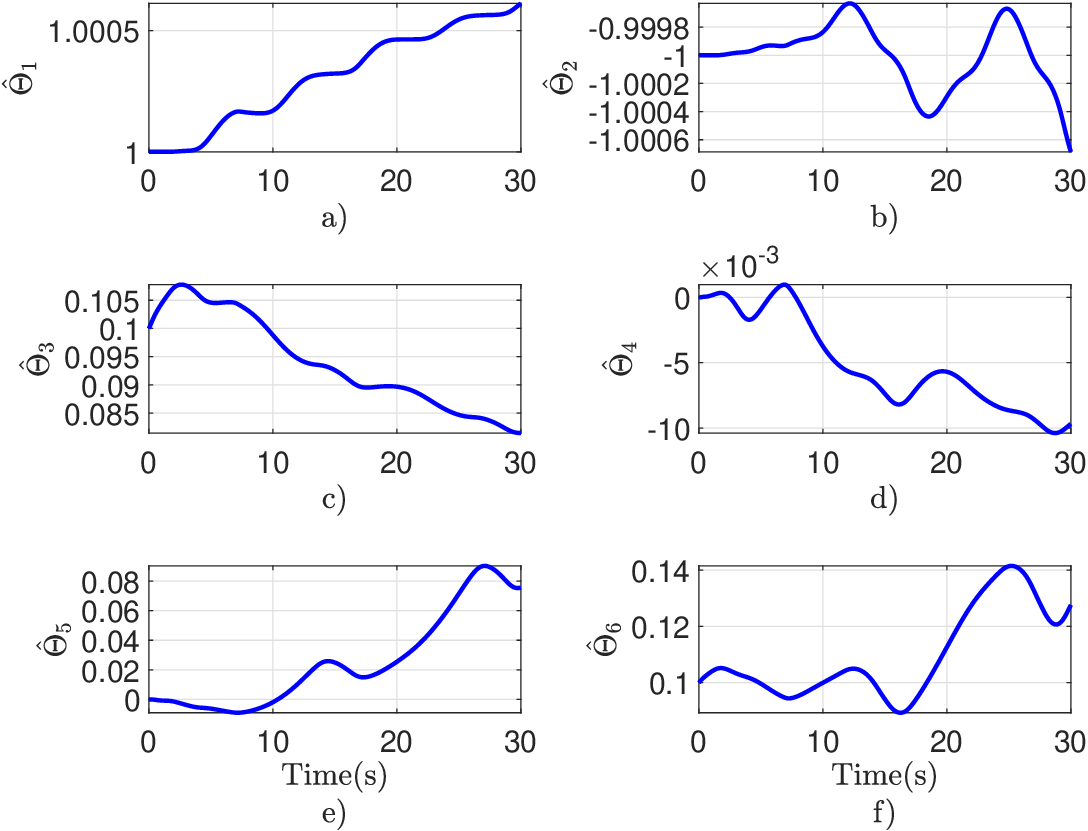}
    \caption{Parameter estimates updated by \eqref{eq:FT_estimator} in the harmonic command-following problem. }
    \label{fig:2dof_iol_adapt_parameters_sinusoidal}
\end{figure}

Figure \ref{fig:2dof_iol_adapt_parameters_sinusoidal} shows the parameter estimates updated by \eqref{eq:FT_estimator}.
Note that the estimates do not necessarily converge to their true values. 

\subsection{Robustness and Sensitivity}
In this section, we investigate the robustness and sensitivity of the adaptive controller to the physical parameters of the system. 
In particular, we vary the system's physical parameters, including moments of inertia, the thrust and torque coefficients, and the thrust directions while keeping the controller and estimator gains fixed. 

\begin{figure}[h]
    \centering
    \includegraphics[width=0.9\columnwidth]{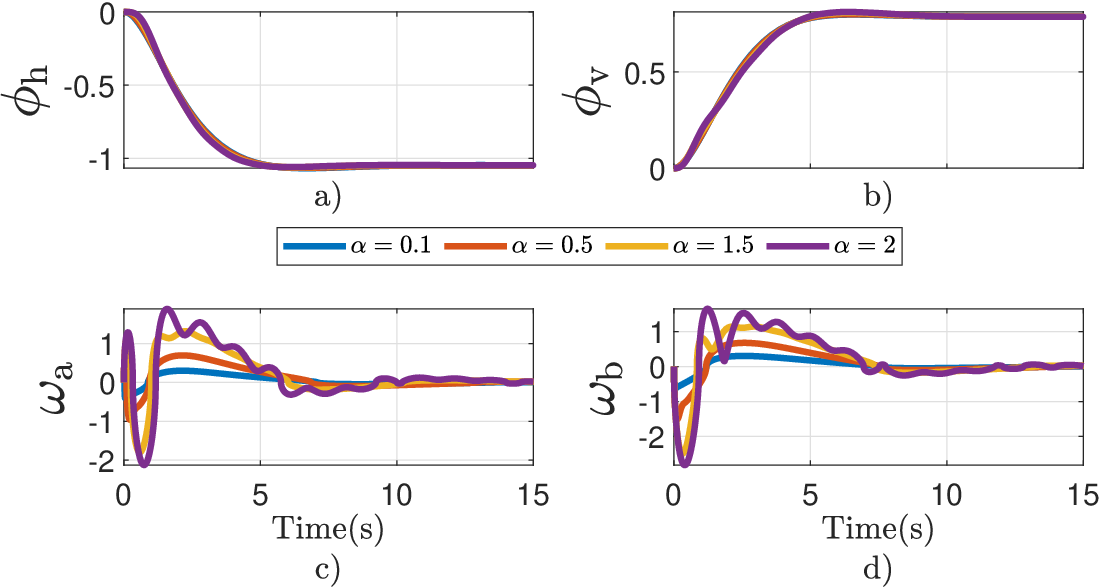}
    \caption{Closed-loop response of the DRRS with the adaptive input-output linearizing controller \eqref{eq:adaptive_IOL_control_law} in the case where moment of inertia is scaled by $\alpha$.}
    \label{fig:sensitivity_inertia}
\end{figure}

First, we vary the moments of inertia by scaling $J_1, J_2, $ and $J_3$ by a scalar $\alpha.$
In particular, we set $\alpha \in \{0.1, 0.5, 1.5, 2\}$. 
Figure \ref{fig:sensitivity_inertia} shows the closed-loop response of the DRRS with scaled moments of inertia, where 
a) and b) show the angles $\phi_\rmv$ and $\phi_\rmh,$ respectively and
c) and d) show the propeller speeds $\omega_\rma$ and $\omega_\rmb,$ respectively.
Note that the angle response remains unaffected by the variation in the moments of inertia, whereas the control required to maintain the closed-loop performance changes significantly. 
This is due to the fact that the adaptive input-output linearizing controller given by \eqref{eq:adaptive_IOL_control_law} results in the same input-output dynamics \eqref{eq:IOlinearizedsystem_x}-\eqref{eq:IOlinearizedsystem_y} irrespective of the parametric values in the physical system. 
The slight change in closed-loop angle response is due to the adaptive nature of the controller \eqref{eq:adaptive_IOL_control_law}.

\begin{figure}[h]
    \centering
    \includegraphics[width=0.9\columnwidth]{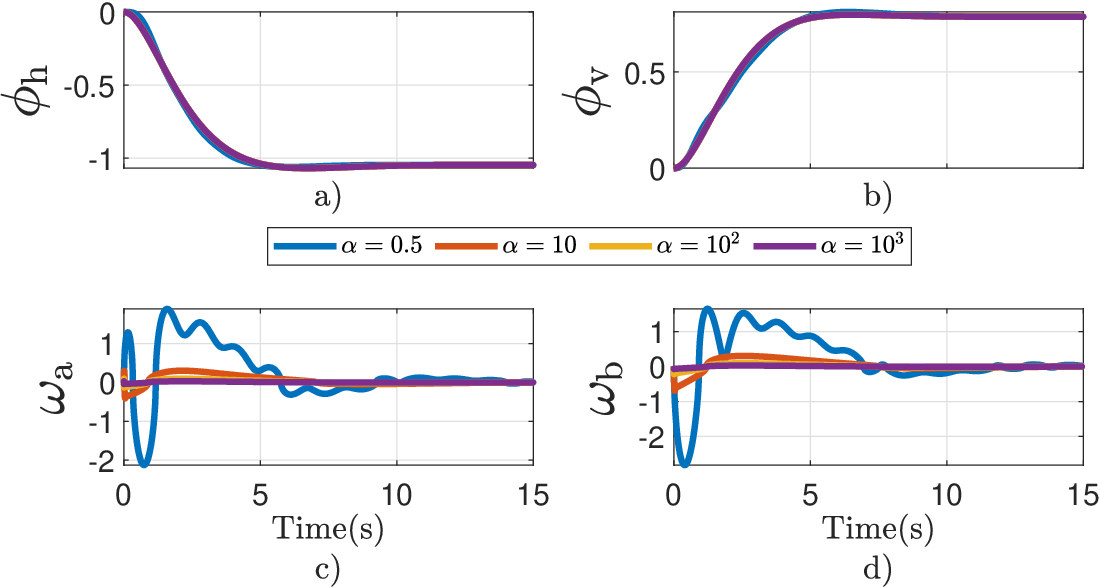}
    \caption{Closed-loop response of the DRRS with the adaptive input-output linearizing controller \eqref{eq:adaptive_IOL_control_law} in the case where thrust and torque coefficients are scaled by $\alpha$.}
    \label{fig:sensitivity_motor}
\end{figure}

Next, we vary the thrust and torque coefficients by scaling $k_\rmf $ and $k_\tau$ by a scalar $\alpha.$
In particular, we set $\alpha \in \{0.5, 10, 10^2, 10^3\}$. 
Figure \ref{fig:sensitivity_motor} shows the closed-loop response of the DRRS with scaled thrust and torque coefficients, where
a) and b) show the angles $\phi_\rmv$ and $\phi_\rmh,$ respectively and
c) and d) show the propeller speeds $\omega_\rma$ and $\omega_\rmb,$ respectively.
Note that, as in the previous case, the angle response remains unaffected by the variations in the thrust and torque coefficients, whereas the control required to maintain the closed-loop performance changes significantly. 
%

\begin{figure}[h]
    \centering
    \includegraphics[width=0.9\columnwidth]{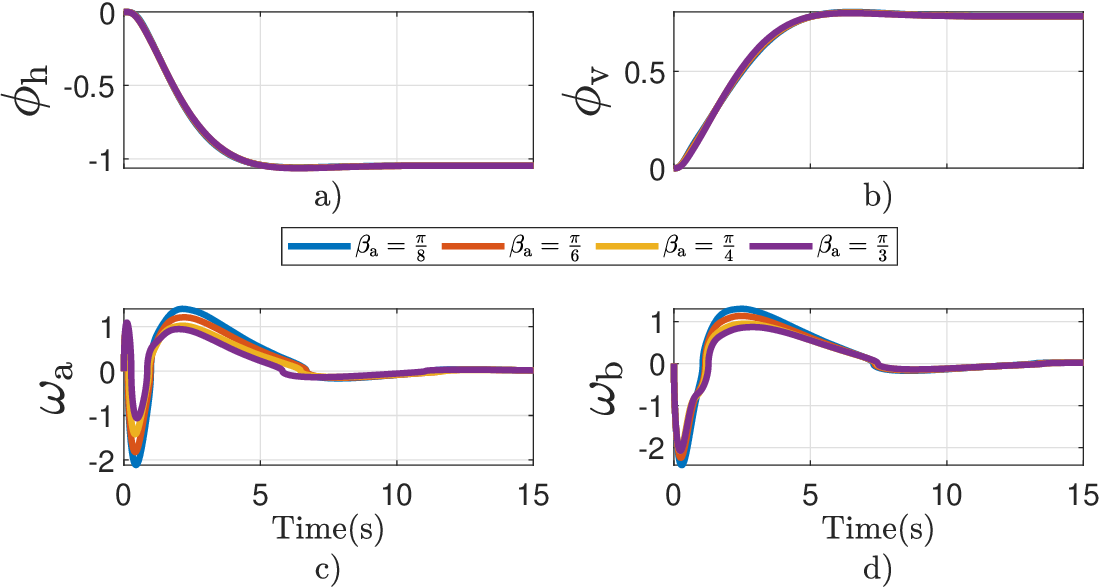}
    \caption{
    Closed-loop response of the DRRS with the adaptive input-output linearizing controller \eqref{eq:adaptive_IOL_control_law} in the case where the rotor axis varies.
    }
    \label{fig:sensitivity_angel}
\end{figure}

Finally, we change the configuration of the DRRS by changing the thrust axis, which is parameterized by $\beta_a$ and $\beta_b.$
In particular, we set $\beta_a \in \{\pi/8, \pi/6, \pi/4, \pi/3\}$ and $\beta_b = -\beta_a.$
Figure \ref{fig:sensitivity_angel} shows the closed-loop response of the DRRS with different motor angle configurations, where
a) and b) show the angles $\phi_\rmv$ and $\phi_\rmh,$ respectively and
c) and d) show the propeller speeds $\omega_\rma$ and $\omega_\rmb,$ respectively.
Note that, as in the previous cases, the angle response remains unaffected by the variations in the system configuration, whereas the control required to maintain the closed-loop performance changes significantly to compensate for the changes in configuration.

\section{Conclusion}
\label{sec:conc}
This paper developed an adaptive controller for a dual-rotor rotational system without requiring prior knowledge of the system's physical parameters. 
The adaptive controller is constructed by combining an input-output linearization scheme with a finite-time convergent parameter estimator, resulting in a composite controller. Additionally, a linear controller is designed for the linearized dynamics to achieve the desired tracking performance.
The effectiveness of the proposed controller is validated through numerical simulations of both constant and harmonic trajectory-tracking problems, and its robustness is demonstrated by varying the system's physical parameters without retuning the controller. 
%

%

A key limitation of the current approach is the singularity of the equations of motion due to using Euler angles.
Our future work is thus focused on developing an adaptive controller that uses direction cosine matrices to parameterize the angular orientation of DRRS to alleviate the singularity problem. 
An alternative approach is to integrate state constraints in the controller to prevent the DRRS from approaching the singularity. 

\bibliography{ifacconf}
\end{document}

%% file: PackagesCommands.tex
\newtheorem{remark}{Remark}
\usepackage{float} 

\usetikzlibrary{shapes,arrows,calc}
\tikzstyle{bigblock} = [draw, fill=blue!20, rectangle, 
    minimum height=6em, minimum width=8em]
\tikzstyle{medblock} = [draw, fill=blue!20, rectangle, 
    minimum height=4em, minimum width=4em]    
\tikzstyle{mux} = [draw, fill=black!20, rectangle, 
    minimum height=5em, minimum width=0.1em]    
\tikzstyle{smallblock} = [draw, fill=blue!20, rectangle, 
    minimum height=2em, minimum width=3em]
    
\tikzstyle{data_block} = [draw, fill=green!20, rectangle, 
    minimum height=2em, minimum width=3em]
\tikzstyle{ops_block} = [draw, fill=blue!20, rectangle, 
    minimum height=2em, minimum width=3em]    
\tikzstyle{est_block} = [draw, fill=red!20, rectangle, 
    minimum height=2em, minimum width=3em]    
    
\tikzstyle{sum} = [draw, fill=blue!20, circle, node distance=1cm,minimum height=0.5cm]
\tikzstyle{signal} = [coordinate]
\tikzstyle{pinstyle} = [pin edge={to-,thin,black}]
\tikzstyle{block} = [draw, fill=blue!20, rectangle, 
    minimum height=3em, minimum width=9em]
\tikzstyle{blockS} = [draw, fill=blue!20, rectangle, 
    minimum height=3em, minimum width=4em]    
\tikzstyle{input} = [coordinate]
\tikzstyle{output} = [coordinate]
\usetikzlibrary{matrix}

\usetikzlibrary{math} 

\usetikzlibrary{patterns}
\usetikzlibrary{calc,patterns,decorations.pathmorphing,decorations.markings}

\newcommand{\bc}{\begin{center}}
\newcommand{\ec}{\end{center}}
\newcommand{\benum}{\begin{enumerate}}
\newcommand{\eenum}{\end{enumerate}}
\newcommand{\nn}{\nonumber}
\newcommand{\matl}{\left[ \begin{array}}
\newcommand{\matr}{\end{array} \right]}

\renewcommand{\matl}{\begin{bmatrix}}
\renewcommand{\matr}{\end{bmatrix}}

\newcommand{\matls}{\left[ \begin{smallmatrix}}
\newcommand{\matrs}{\end{smallmatrix} \right]}
\newcommand{\isdef}{\stackrel{\triangle}{=}}

\newcommand{\inv}{^{-1}}

\newcommand{\vect}[1]{\overset{\rightharpoonup}{#1}}

\newcommand{\rmA}{{\rm A}}
\newcommand{\rmB}{{\rm B}}
\newcommand{\rmC}{{\rm C}}

\newcommand{\rmF}{{\rm F}}

\newcommand{\rmT}{{\rm T}}

\newcommand{\rma}{{\rm a}}
\newcommand{\rmb}{{\rm b}}
\newcommand{\rmc}{{\rm c}}
\newcommand{\rmd}{{\rm d}}

\newcommand{\rmf}{{\rm f}}

\newcommand{\rmh}{{\rm h}}

\newcommand{\rmr}{{\rm r}}

\newcommand{\rmv}{{\rm v}}

\newcommand{\BBN}{{\mathbb N}}

\newcommand{\BBR}{{\mathbb R}}

\newcommand{\SB}{{\mathcal B}}
\newcommand{\SC}{{\mathcal C}}

\newcommand{\SF}{{\mathcal F}}
\newcommand{\SG}{{\mathcal G}}

\newcommand{\neweqline}{\ensuremath{\nn \\ &\quad }}

\newcommand{\framedot}[2]{\stackrel{{\rm #1}\bullet}{#2}}

\newcommand{\tarrow}[1]{\overset{\rightarrow}{#1}}

\usepackage{enumitem,amssymb}
\usepackage{pifont}
\newcommand{\ihat}[1]{\ensuremath{\hat \imath_{\rm #1} }}
\newcommand{\jhat}[1]{\ensuremath{\hat \jmath_{\rm #1} }}
\newcommand{\khat}[1]{\ensuremath{\hat k_{\rm #1} }}

%% file: main_2025_03_10.bbl
\begin{thebibliography}{13}
\providecommand{\natexlab}[1]{#1}
\providecommand{\url}[1]{\texttt{#1}}
\providecommand{\urlprefix}{URL }
\expandafter\ifx\csname urlstyle\endcsname\relax
  \providecommand{\doi}[1]{doi:\discretionary{}{}{}#1}\else
  \providecommand{\doi}{doi:\discretionary{}{}{}\begingroup
  \urlstyle{rm}\Url}\fi

\bibitem[{AlHamouch et~al.(2019)AlHamouch, Tuqan, Bardawil, and
  Daher}]{Investigating}
AlHamouch, A., Tuqan, M., Bardawil, C., and Daher, N. (2019).
\newblock Investigating performance of adaptive and robust control schemes for
  quanser aero.
\newblock In \emph{2019 Fourth International Conference on Advances in
  Computational Tools for Engineering Applications (ACTEA)}, 1--6.
\newblock \doi{10.1109/ACTEA.2019.8851102}.

\bibitem[{Baciu and Lazar(2024)}]{baciu2024model}
Baciu, A. and Lazar, C. (2024).
\newblock Model-free adaptive pitch control for a nonlinear aerospace
  laboratory equipment.
\newblock In \emph{2024 28th International Conference on System Theory, Control
  and Computing (ICSTCC)}, 212--216. IEEE.

\bibitem[{Dyvik et~al.(2023)Dyvik, Fjereide, and Rotondo}]{dyvik2023modeling}
Dyvik, M., Fjereide, D.E., and Rotondo, D. (2023).
\newblock Modeling and identification of the quanser aero using a detailed
  description of friction and centripetal forces.
\newblock In \emph{Proceedings of the 64th International Conference of
  Scandinavian Simulation Society}. Link{\"o}ping Universitet.

\bibitem[{Fellag and Belhocine(2024)}]{fellag20242}
Fellag, R. and Belhocine, M. (2024).
\newblock 2-dof helicopter control via state feedback and full/reduced-order
  observers.
\newblock In \emph{2024 2nd International Conference on Electrical Engineering
  and Automatic Control (ICEEAC)}, 1--6. IEEE.

\bibitem[{Frasik and Gabrielsen(2018)}]{frasik2018practical}
Frasik, J.M. and Gabrielsen, S.I.L. (2018).
\newblock \emph{Practical application of advanced control: An evaluation of
  control methods on a quanser aero}.
\newblock Master's thesis, Universitetet i Agder; University of Agder.

\bibitem[{Labdai et~al.(2020)Labdai, Chrifi-Alaoui, Drid, Delahoche, and
  Bussy}]{adaptive-slide}
Labdai, S., Chrifi-Alaoui, L., Drid, S., Delahoche, L., and Bussy, P. (2020).
\newblock Real-time implementation of an optimized fractional sliding mode
  controller on the quanser-aero helicopter.
\newblock In \emph{2020 International Conference on Control, Automation and
  Diagnosis (ICCAD)}, 1--6.
\newblock \doi{10.1109/ICCAD49821.2020.9260546}.

\bibitem[{Pereda~P{\'e}rez(2024)}]{pereda2024modeling}
Pereda~P{\'e}rez, G. (2024).
\newblock \emph{Modeling and control using feedback linearization of a Quanser
  Aero 2 device}.
\newblock Master's thesis, Universitat Polit{\`e}cnica de Catalunya.

\bibitem[{Portella~Delgado and Goel(2024{\natexlab{a}})}]{portella2024adaptive}
Portella~Delgado, J.M. and Goel, A. (2024{\natexlab{a}}).
\newblock Adaptive nonlinear control of a bicopter with unknown dynamics.
\newblock In \emph{2024 American Control Conference (ACC)}, 3867--3872. IEEE.

\bibitem[{Portella~Delgado and
  Goel(2024{\natexlab{b}})}]{portella2024circumventing}
Portella~Delgado, J.M. and Goel, A. (2024{\natexlab{b}}).
\newblock Circumventing unstable zero dynamics in input-output linearization of
  longitudinal flight dynamics.
\newblock In \emph{AIAA SCITECH 2024 Forum}, 321.

\bibitem[{Schlanbusch(2019)}]{schlanbusch2019adaptive}
Schlanbusch, S.M. (2019).
\newblock \emph{Adaptive backstepping control of quanser 2DOF helicopter:
  Theory and experiments}.
\newblock Master's thesis, Universitetet i Agder; University of Agder.

\bibitem[{Schlanbusch and Zhou(2024)}]{schlanbusch2024adaptive}
Schlanbusch, S.M. and Zhou, J. (2024).
\newblock Adaptive predictor-based control for a helicopter system with input
  delays: Design and experiments.
\newblock \emph{Journal of Automation and Intelligence}, 3(1), 50--56.

\bibitem[{Schäfer et~al.(2024)Schäfer, Rehrl, Huber, and Hirlaender}]{id}
Schäfer, G., Rehrl, J., Huber, S., and Hirlaender, S. (2024).
\newblock Comparison of model predictive control and proximal policy
  optimization for a 1-dof helicopter system.
\newblock In \emph{2024 IEEE 22nd International Conference on Industrial
  Informatics (INDIN)}, 1--7.
\newblock \doi{10.1109/INDIN58382.2024.10774357}.

\bibitem[{Steinbusch and Reyhanoglu(2019)}]{back2}
Steinbusch, A. and Reyhanoglu, M. (2019).
\newblock Robust nonlinear tracking control of a 2-dof helicopter system.
\newblock In \emph{2019 12th Asian Control Conference (ASCC)}, 1649--1654.

\end{thebibliography}
